\documentclass[aps,prb,twocolumn,showpacs,amsmath,amssymb,superscriptaddress]{revtex4}
\usepackage{graphicx}
\usepackage{amssymb}
\usepackage{natbib}
\usepackage{color}

\usepackage{epstopdf}

\newcommand{\beq}{\begin{eqnarray}}
\newcommand{\eeq}{\end{eqnarray}}

\begin{document}

\title{Orbital selective neutron spin resonance in underdoped superconducting NaFe$_\textbf{0.985}$Co$_\textbf{0.015}$As}

\author{Weiyi Wang}
\affiliation{Department of Physics and Astronomy, Rice University, Houston, Texas 77005, USA}
\author{J. T. Park}
\affiliation{Heinz Maier-Leibnitz Zentrum (MLZ), Technische Universit$\ddot{a}$t M$\ddot{u}$nchen, D-85747 Garching, Germany}
\author{Rong Yu}
\affiliation{Department of Physics and Beijing Key Laboratory of Opto-electronic Functional Materials and Micro-nano Devices, Renmin University of China, Beijing 100872, China}
\affiliation{Department of Physics and Astronomy, Shanghai Jiao Tong University, Shanghai 200240, China and Collaborative Innovation Center for Advanced Microstructures, Nanjing 210093, China}
\author{Yu Li}
\affiliation{Department of Physics and Astronomy, Rice University, Houston, Texas 77005, USA}
\author{Yu Song}
\affiliation{Department of Physics and Astronomy, Rice University, Houston, Texas 77005, USA}
\author{Zongyuan Zhang}
\affiliation{School of Physics, Huazhong University of Science and Technology, Wuhan, Hubei 430074, China }
\author{Alexandre Ivanov}
\affiliation{Institut Laue-Langevin, 71, avenue des Martyrs, 38000 Grenoble, France}
\author{Jiri Kulda}
\affiliation{Institut Laue-Langevin, 71, avenue des Martyrs, 38000 Grenoble, France}
\author{Pengcheng Dai}
\email{pdai@rice.edu}
\affiliation{Department of Physics and Astronomy, Rice University, Houston, Texas 77005, USA}
\affiliation{Center for Advanced Quantum Studies and Department of Physics, Beijing Normal University, Beijing 100875, China}

\begin{abstract}
We use neutron scattering to study the electron-doped superconducting 
NaFe$_{0.985}$Co$_{0.015}$As ($T_c=14$ K), which has co-existing static antiferromagnetic (AF) order ($T_N=31$ K) and exhibits two 
neutron spin resonances ($E_{r1}\approx 3.5$ meV and $E_{r2}\approx 6$ meV)
at the in-plane AF ordering wave vector ${\bf Q}_{\rm AF}={\bf Q}_{1}=(1,0)$ in reciprocal space. In the twinned state below the
tetragonal-to-orthorhombic structural transition $T_s$, 
both resonance modes appear at ${\bf Q}_{1}$ but cannot be distinguished from ${\bf Q}_{2}=(0,1)$.  By detwinning the single crystal with uniaxial pressure along the orthorhombic $b$-axis, we find that both resonances appear only at ${\bf Q}_{1}$ with vanishing intensity at ${\bf Q}_{2}$. 
Since electronic bands of the orbital $d_{xz}$ and $d_{yz}$ characters split below $T_s$ 
with the $d_{xz}$ band sinking $\sim10$ meV below the Fermi surface, our results indicate that 
the neutron spin resonances in NaFe$_{0.985}$Co$_{0.015}$As arise mostly from  
quasi-particle excitations between the hole and electron Fermi surfaces with the $d_{yz}$ orbital character. 
\end{abstract}


\maketitle

Understanding the role of magnetism in the electron pairing of unconventional superconductors such as 
copper oxides, iron pnictides, and heavy Fermions continues to be an important topic in modern condensed matter physics because superconductivity in these materials is derived from their long-range antiferromagnetic (AF) ordered parent compounds \cite{scalapino,BKeimer,P.Dai,frank}.  One of the key evidences suggesting that magnetism is involved in the electron pairing and superconductivity is the observation by inelastic neutron scattering (INS) of a neutron spin resonance in the superconducting state of 
various unconventional superconductors \cite{J.Rossat-Mignod,eschrig,christianson,lumsden,schi09,dsinosov09,clzhang13,C.Zhang,Zhang2016,C.Stock,Raymond,song16}.  The resonance is a collective magnetic excitation occurring below the superconducting 
transition temperature $T_c$ with a temperature-dependence similar to the superconducting order parameter, and is 
located at the AF ordering wave vector ${\bf Q}_{\rm AF}$ of their parent compound \cite{J.Rossat-Mignod,eschrig,christianson,lumsden,schi09,dsinosov09,clzhang13,C.Zhang,Zhang2016,C.Stock,Raymond,song16}. 
Moreover, the energy of the resonance has been associated with $T_c$ or superconducting gap size $\Delta$ \cite{D.Inosov,G.Yu}. 
In iron pnictide superconductors [Fig. 1(a)], the resonance is generally interpreted as a spin exciton arising 
from sign-reversed quasiparticle excitations between the hole (at $\Gamma$ point) and electron (at $X$ and $Y$ points) Fermi 
surfaces [Fig. 1(b)] \cite{hirschfeld,chubukov}. In reciprocal space, the $\Gamma-X$ and $\Gamma-Y$ Fermi surface nesting corresponds to
wave vectors of ${\bf Q}_1=(1,0)$ and ${\bf Q}_2=(0,1)$, respectively [Fig. 1(c)].
If this is indeed the case, one would expect that significant modifications of the Fermi surfaces should affect the wave vector dependence and energy of the resonance \cite{P.Dai}. 

In electron-doped superconducting NaFe$_{1-x}$Co$_{x}$As [Fig. 1(a)] \cite{Parker10,S.Li,G.Tan,Tan16}, INS 
experiments have 
mapped out the Co-doping dependence of the resonance \cite{clzhang13,C.Zhang,Zhang2016}. 
For underdoped NaFe$_{0.985}$Co$_{0.015}$As with $T_c=14$ K and a tetragonal-to-orthorhombic 
structural transition below $T_s\approx 40$ K, where 
the static long-range AF order ($T_N=31$ K) microscopically coexists with superconductivity \cite{C.Zhang,P.Cai,Oh13,Ma14}, 
superconductivity induces one dispersive sharp resonance near E$_{r1}$ = 3.5 meV and a broad dispersionless mode at $E_{r2}$ = 6 meV at
a wave vector consistent with ${\bf Q}_{\rm AF}={\bf Q}_1=(1,0)$ but cannot be distinguished from  
${\bf Q}_2=(0,1)$ \cite{C.Zhang}. Although nuclear magnetic resonance (NMR) measurements on NaFe$_{1-x}$Co$_{x}$As  
suggested the presence of a 8\% volume fraction paramagnetic phase for $x<0.0175$ samples that is doping independent \cite{Ma14}, 
the bulk nature of neutron scattering does not allow us to separate this phase from the dominant AF phase. 
Upon further Co-doping to 
reach optimal superconductivity without static AF order, 
the double resonances in NaFe$_{1-x}$Co$_{x}$As become a single resonance \cite{clzhang13,C.Zhang,Zhang2016}.  
Since the disappearance of the double resonances occurs 
at approximately the same doping level as the vanishing static AF order with increasing Co-doping \cite{clzhang13,C.Zhang,Zhang2016}, the 
presence of double resonances has been interpreted as due to the coexisting AF order and superconductivity \cite{Knolle2011,Lv2014}.  
In this picture, one would expect that the resonance associated with the AF order to exclusively appear 
at ${\bf Q}_{\rm AF}={\bf Q}_1=(1,0)$ in a completely detwinned sample as the collinear AF order explicitly breaks the $C_4$ rotational 
symmetry of the orthorhombic lattice [see inset of Fig. 1(a)], while the resonance associated with itinerant electrons and 
simple nested Fermi surfaces (without considering the inter- and intra- orbital scattering processes)
should be present at both  ${\bf Q}_{\rm AF}={\bf Q}_1$ and ${\bf Q}={\bf Q}_2$ [Fig. 1(b)] \cite{Knolle2011,Lv2014}. 

\begin{figure}[ht]
	\centering
	\includegraphics[scale = 0.4]{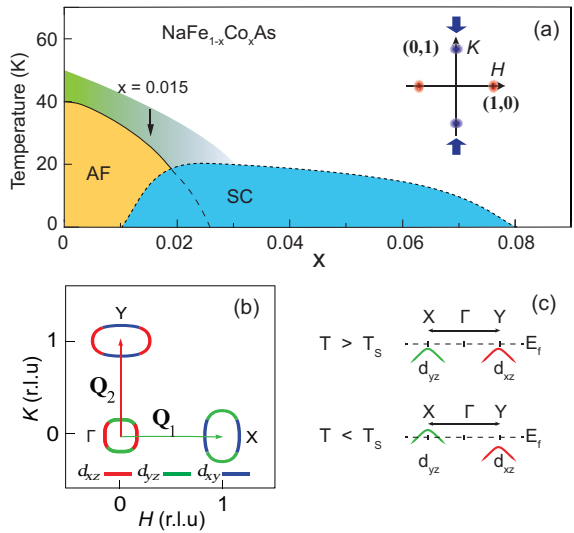}
	\caption{(a) The phase diagram of NaFe$_{1-x}$Co$_{x}$As with the arrow indicating the Co-doping concentration in our experiment. 
	The inset shows positions of magnetic excitations in the $[H,K]$ plane under uniaxial pressure along the $b$-axis. 
	(b) Schematic Fermi surfaces of NaFe$_{0.985}$Co$_{0.015}$As in the paramagnetic tetragonal state and different orbitals are characterized with different colors. The arrows mark nesting wave vectors $\textbf{Q}_\textbf{1}$ = (1,0) and $\textbf{Q}_\textbf{2}$ = (0,1). 
	(c) Schematic $d_{yz}$ and $d_{xz}$ orbital bands in NaFe$_{1-x}$Co$_{x}$As above and below $T_s$ as seen by ARPES \cite{M.Yi,Y.Zhang}. }
	\label{fig:Figure1}
\end{figure}

Alternatively, the presence of double resonances can arise from orbital-selective pairing-induced superconducting gap anisotropy \cite{R.Yu}.
From angle resolved photoemission spectroscopy (ARPES) experiments \cite{Y.Zhang,M.Yi,Q.Ge}, it was found that the superconducting gap anisotropy appearing in the low Co-doping regime of NaFe$_{1-x}$Co$_{x}$As disappears in electron overdoped NaFe$_{0.955}$Co$_{0.045}$As \cite{Z.Liu,S.Thirupathaiah}. The double resonances 
at ${\bf Q}_{\rm AF}={\bf Q}_1$ in underdoped NaFe$_{1-x}$Co$_{x}$As 
can therefore be due to  the presence of 
superconducting gap anisotropy in the underdoped regime \cite{C.Zhang}. Since AF order is not expected to affect the superconducting gap anisotropy,  
one would expect the presence of the double resonances at ${\bf Q}_{\rm AF}={\bf Q}_1$ and ${\bf Q}={\bf Q}_2$ in a 
detwinned single crystal of NaFe$_{0.985}$Co$_{0.015}$As \cite{R.Yu}.  Therefore, by using uniaxial pressure to 
detwin the single crystal \cite{dhital,Dhital14,YSong13,X.Lu}, one can potentially determine the microscopic origin of the 
double resonances \cite{C.Zhang3}.
In previous INS experiment on partially detwinned NaFe$_{0.985}$Co$_{0.015}$As, it was reported that the double resonance
are present with similar intensity 
at both ${\bf Q}_{\rm AF}={\bf Q}_1$ and ${\bf Q}={\bf Q}_2$, thus suggesting that the double resonance originates from the anisotropic superconducting gap in the underdoped regime \cite{C.Zhang3}. However, the detwinning ratio of studied compound was estimated from two separate experiments under possibly not identical pressure condition.  In addition, due to the large background level at the elastic scattering channel originated from the pressure device in the experiment, the reported detwinning ratio in \cite{C.Zhang3} is likely to be overestimated. 
Therefore, to conclusively determine the effect of detwinning and uniaxial pressure on the resonance, one needs to carry out INS experiments
by comparing directly pressured and pressureless case using the same sample holder with the same spectrometer setup.

 If we assume that low energy spin excitations in iron pnictides 
originate from quasi-particle excitations between hole and electron Fermi surfaces [Fig. 1(b)], 
the orbital characters of hole and electron Fermi surfaces should determine the nature of observed 
spin excitations at ${\bf Q}_1$ and ${\bf Q}_2$ \cite{hirschfeld,chubukov}. For example, INS experiments on
 LiFe$_{1-x}$Co$_{x}$As system reveal that transverse incommensurate spin excitations observed in superconducting LiFeAs \cite{Qureshi12,Mwang12} 
change to commensurate spin excitations for nonsuperconducting LiFe$_{0.88}$Co$_{0.12}$As arising mostly from the hole-electron Fermi surface
nesting of the d$_{xy}$ orbitals, thus suggesting that Fermi surface nesting conditions of the $d_{yz}$ and $d_{xz}$ orbitals are important for superconductivity \cite{L.Yu}. In the case of NaFe$_{1-x}$Co$_{x}$As \cite{Parker10,S.Li,G.Tan,Tan16}, ARPES measurements 
on uniaxial pressure detwinned single crystals reveal the splitting of the $d_{xz}$ and $d_{yz}$ orbitals at temperatures below $T_s$ (although in case of large pressure, the splitting actually first takes place at temperatures above $T_s$), where
the bands of dominant $d_{yz}$ orbital character shift up in $\Gamma-X$ direction (${\bf Q}_1$) and bands of dominant $d_{xz}$ orbital character sink below Fermi surface in $\Gamma-Y$ direction (${\bf Q}_2$) [Figure 1(c)] \cite{M.Yi,Y.Zhang}.  This means that low-energy spin excitations  
at wave vectors ${\bf Q}_1$ and ${\bf Q}_2$ should behave differently in the low-temperature 
superconducting state. Since bands of dominant $d_{yz}$ orbital characters sink below Fermi surface below $T_s$,
neutron spin resonance associated with quasiparticle excitations of hole-electron Fermi surfaces of the $d_{yz}$ orbitals at ${\bf Q}_2$
should be absent below $T_c$, while the resonance associated with $d_{yz}$ and $d_{xy}$ orbitals should appear below $T_c$ at
${\bf Q}_1$  [Figs. 1(b) and 1(c)].

To test if this is indeed the case, we carried out INS experiments on uniaxial detwinned NaFe$_{0.985}$Co$_{0.015}$As single crystal. 
Compared with earlier experiments on the same doping concentration \cite{C.Zhang3}, the new measurements have much better statistics and collected uniaxial pressured/pressureless data using the same experimental setup. We find the presence of double resonance at ${\bf Q}_1$ and ${\bf Q}_2$ with intensity ratio of the modes between 
 ${\bf Q}_{1}$ and ${\bf Q}_{2}$ agreeing well with the detwinning ratio obtained using magnetic Bragg peaks at these two wave vectors.
These results therefore indicate that superconductivity induced resonance arises from  the nesting of hole-electron Fermi surfaces with 
dominant $d_{yz}$ orbital characters.

\begin{figure}[t]
	\includegraphics[scale = 0.4]{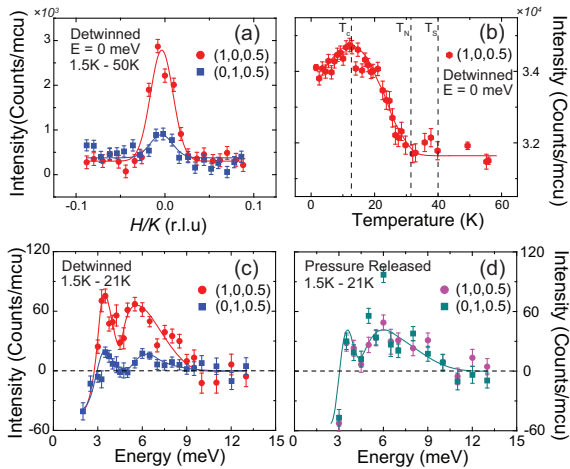}
	\centering
	\caption{ (a) Temperature differences of transverse scans at the $(1,0,0.5)$ and $(0,1,0.5)$ magnetic Bragg peak positions in NaFe$_{0.985}$Co$_{0.015}$As single crystal under uniaxial pressure of {\raise.17ex\hbox{$\scriptstyle\mathtt{\sim}$}}10 MPa. (b) Temperature dependence of the AF peak intensity at ${\bf Q}_{1} = (1,0,0.5)$ under uniaxial pressure with vertical dashed lines indicating $T_c = 14$ K, $T_N = 31$ K, 
	and $T_s = 40$ K. (c) Neutron spin resonance modes at ${\bf Q}_{1} =(1,0,0.5)$ and ${\bf Q}_{2} =(0,1,0.5)$ are obtained by taking temperature differences of the constant-${\bf Q}$ scans below and above $T_c$ in uniaxial pressure partially detwinned sample. (d) Neutron spin resonance modes in pressure released sample obtained using the same sample and the same sample holder with the same spectrometer setup
	as in (c). The similar intensity of the resonance at ${\bf Q}_{1}$ and ${\bf Q}_{1}$ indicate that the sample becomes nearly 100\% twinned.
	}
	\label{fig:Figure2}
\end{figure}

Our neutron scattering experiment was carried out on IN8-Thermal neutron three-axis spectrometer at Institut Laue-Langevin, Grenoble, France. We used horizontally and vertically focused pyrolytic graphite [PG(002)] monochromator and analyzer with fixed scattered (final) energy 
$E_f = 14.68$ meV.  The high order harmonics from the PG(002) monochromator are suppressed by an oriented PG-filter 
in the scattered beam. 
Using structural orthorhombic unit cell with lattice parameters $a\approx b \approx 5.5968 \text{\AA}$ and $ c \approx 6.9561 \text{\AA}$ at $T = 1.5$ K, we denote the momentum transfer $\textbf{Q}=H\textbf{a}^\ast+K\textbf{b}^\ast+L\textbf{c}^\ast$ 
as $\textbf{Q}=(H,K,L)$ in reciprocal lattice units (r.l.u.) with ${\bf a}^\ast = \hat{{\bf a}}2\pi/a$, ${\bf b}^\ast=\hat{{\bf b}}2\pi/b$ and ${\bf c}^\ast=\hat{{\bf c}}2\pi/c$.  In the AF ordered state of a completely detwinned sample with uniaxial pressure applied along
the $b$-axis direction, AF Bragg peaks occur at $\textbf{Q}=(1,0,L)$ with $L = 0.5,1.5,2.5\ldots$ and there are no magnetic peaks at $(0,1,L)$ \cite{S.Li}.

High-quality NaFe$_{0.985}$Co$_{0.015}$As single crystals are prepared by self-flux method \cite{M.Tanatar}, and we cut one large single crystal ({\raise.17ex\hbox{$\scriptstyle\mathtt{\sim}$}}300 mg) into the rectangular shape along the $[1,0,0]$ and $[0,1,0]$ directions. The sample is mounted inside aluminum-based sample holder with a uniaxial pressure of $P\approx 10$ MPa along the $b$-axis direction (although it is rather difficult to precisely determine the magnitude of the actual uniaxial strain on the sample). Similar to previous neutron works \cite{X.Lu}, we align the sample in the $[1,0,0.5]\times[0,1,0.5]$ scattering plane. In such a scattering geometry, we are able to measure the static magnetic order and excitations at both ${\bf Q}_{1}$ = $(1,0,0.5)$ and ${\bf Q}_{2}$ = $(0,1,0.5)$.

Figure 2(a) shows background subtracted elastic transverse scans across ${\bf Q}_{1}$ and ${\bf Q}_{2}$ \cite{SI}. By comparing the intensities between these two positions, we estimate that the sample has a detwinning ratio $\eta=[I(1,0)-I(0,1)]/[I(1,0)+I(0,1)]\approx 62.4\% $. 
Temperature dependence of the magnetic order parameter measured at ${\bf Q}_{1}$
reveals $T_N\approx 31$ K and a suppression of the static AF order below $T_c\approx 14$ K [Fig. 2(b)]. Figure 2(c) shows temperature difference of the energy scans at ${\bf Q}_{1}$ and ${\bf Q}_{2}$ below and above $T_c$. Similar to the results in twinned sample \cite{C.Zhang}, two neutron spin resonance modes are found at 
$E_{r1}\approx 3.5$ meV and $E_{r2}\approx 6$ meV, respectively, and a spin gap opens below $E=3$ meV in the superconducting state. 
Moreover, intensities for both resonance modes are higher at the AF position ${\bf Q}_{1}$ than at ${\bf Q}_{2}$.  This is different 
from our previous data obtained on PUMA \cite{C.Zhang3}, which is likely due to 
the improved detwinning ratio in the present study. To further confirm that such difference is induced by uniaxial strain, we released the uniaxial pressure and carried out same energy scans on the same sample under same experiment setup. Figure 2(d) shows temperature difference of the energy scans upon releasing the uniaxial pressure. 
As expected, the sample goes back to the twinned state, and there are no observable 
differences of the both resonance modes between ${\bf Q}_{1}$ and ${\bf Q}_{2}$.

Since our experiments are carried out using the same spectrometer setup 
on the same unaxial pressure detwinned and twinned (pressure released) sample with the same sample holder, we are able to compare 
the effect of uniaxial pressure on the double resonances directly. Figure 3(a) shows the intensity sum of the double resonances at ${\bf Q}_{1}$ and ${\bf Q}_{2}$ in partially detwinned sample and twinned sample. 
We find that the resonance intensities are identical in these two cases, thus indicating that the intensity gain at ${\bf Q}_{1}$ in detwinned sample comes from the intensity loss at ${\bf Q}_{2}$. Figure 3(b) shows the ratio of resonance intensities between ${\bf Q}_{1}$ and ${\bf Q}_{2}$ in the partially detwinned sample. Since the intensity ratios for both resonance modes agree well with the detwinning ratio obtained using magnetic Bragg peaks [Fig. 2(a)], we conclude that neutron spin resonance modes in this system only appear at ${\bf Q}_{1}$ in a 100\% detwinned sample. 

\begin{figure}[t]
	\centering
	\includegraphics[scale = 0.4]{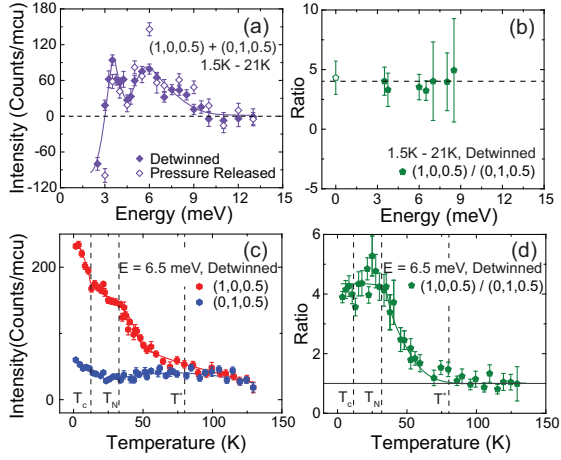}
	\caption{(a) Sum of neutron spin resonance mode intensities at ${\bf Q}_{1}$ and ${\bf Q}_{2}$ in partially detwinned and pressure released sample separately. (b) The ratio between temperature difference of constant-${\bf Q}$ scans (below and above $T_c$) at ${\bf Q}_{1}$ and ${\bf Q}_{2}$ in partially detwinned sample. Points with large error bars around $E=4.5$ meV are not shown. The open circle labels the detwinning ratio measured from the static AF order. (c) Temperature dependence of background subtracted 
	spin excitations at $E_{r2} = 6.5$ meV at ${\bf Q}_{1}$ and ${\bf Q}_{2}$ in the partially 
	detwinned sample \cite{SI}. (d) The corresponding ratio of temperature dependence of spin excitations at $E_{r2} = 6.5$ meV between ${\bf Q}_{1}$ and ${\bf Q}_{2}$. $T_{c}$, $T_{N}$ and $T^{*}>T_s$ are labeled with dashed lines in (c) and (d).}
	\label{fig:Figure3}
\end{figure}

Figure 3(c) shows background subtracted temperature dependence of the second resonance ($E_{r2}$ = 6.5 meV) measured at 
${\bf Q}_{1}$ and ${\bf Q}_{2}$ under uniaxial pressure \cite{SI}. The intensity kink at $T_{N}$ and strong increase below $T_{c}$ at both wave vectors 
agree with previous INS results in twinned sample\cite{C.Zhang}. 
At temperatures well above $T_c$, $T_N$, and $T_s$, magnetic scattering at ${\bf Q}_{1}$ and ${\bf Q}_{2}$ are identical and independent of
the applied uniaxial pressure.  On cooling to 80 K ($>T_s$), we start to see higher magnetic scattering at ${\bf Q}_{1}$, consistent with earlier 
work on other iron pnictide superconductors suggesting the presence of a spin nematic phase \cite{X.Lu,Y.Song}. The intensity ratio between ${\bf Q}_{1}$ and ${\bf Q}_{2}$ at $E_{r2}$ = 6.5 meV is shown in Fig. 3(d). The clear spin excitation anisotropy above $T_s$ is likely due to the 
	applied uniaxial pressure as discussed in Refs. \cite{man15,xlu16}.
As a function of decreasing temperature, the magnetic scattering  
 anisotropy starts to build up below T$^\ast$, saturates at temperatures slightly below $T_{N}$, and shows no anomaly across $T_c$.

\begin{figure}[t]
	\centering
	\includegraphics[scale = 0.45]{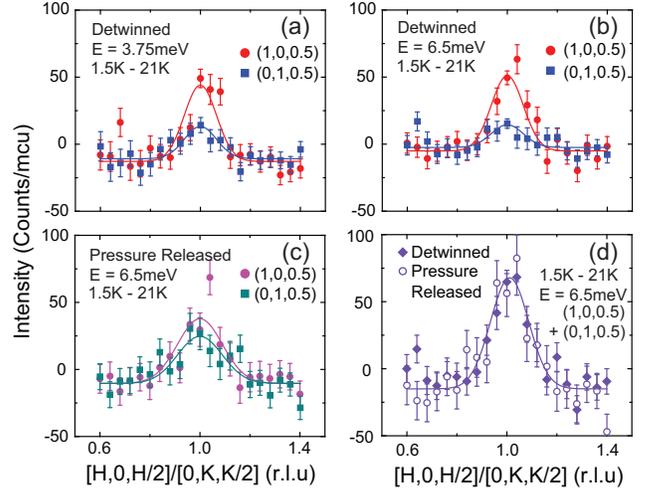}
	\caption{ Temperature differences of constant energy scans at ${\bf Q}_{1}$ and ${\bf Q}_{2}$ between 1.5 K and 21 K at (a)
	$E_{r1}=3.75$ meV, (b) $E_{r2}=6.5$ meV in partially detwinned sample, and (c) $E_{r2}=6.5$ meV in pressure released sample. (d) Comparison
	of sum of the magnetic scattering at ${\bf Q}_{1}$ and ${\bf Q}_{2}$ for partially detwinned and twinned sample.}
	\label{fig:Figure4}
\end{figure}

Figures 4(a) and 4(b) summarize the wave vector dependence of the double resonances in the partially detwinned sample at ${\bf Q}_{1}$ and ${\bf Q}_{2}$. 
The intensity ratios between ${\bf Q}_{1}$ and ${\bf Q}_{2}$ at both resonance energies E$_{r1} \approx 3.75$ meV and E$_{r2} \approx 6.5$ meV are consistent with the detwinning ratio at elastic position. 
When the uniaxial pressure is released, we find no difference between 
${\bf Q}_{1}$ and ${\bf Q}_{2}$ at the resonance energy and the sample goes back to the twinned state [Fig. 4(c)].
Figure 4(d) compares the sum of the resonance intensity at ${\bf Q}_{1}$ and ${\bf Q}_{2}$ for pressured (solid diamonds) and
pressure free case (open circles). To within the statistics of our measurements, we find them to be identical.

Several different theories have been proposed to understand the double resonances \cite{Knolle2011,Lv2014,R.Yu}. In the theory
of coexisting static AF order and superconductivity \cite{Knolle2011,Lv2014}, the AF order leads to a reconstruction of the Fermi surface, which gives rise to different resonance energies $E_{r1}$ and $E_{r2}$ at wave vectors $\mathbf{Q}_1$ and $\mathbf{Q}_2$, respectively. In a twinned sample, double resonances should appear at both $\mathbf{Q}_1$ and $\mathbf{Q}_2$. 
Since $E_{r1}$ is expected to be related with the static AF order and its associated spin waves, it should only appear at the AF ordering wave vector  ${\bf Q}_{1}$ in a detwinned sample, while $E_{r2}$ associated with Fermi surface nesting should appear at both 
$\mathbf{Q}_1$ and $\mathbf{Q}_2$ \cite{Knolle2011,Lv2014}. However, the simultaneously enhancement (suppression) of both resonance peaks at $\mathbf{Q}_1$ ($\mathbf{Q}_2$) observed in our experiment for a detwinned sample appears to rule out this theory. This is consistent with polarized inelastic neutron 
scattering experiments, which reveal that low-energy spin waves ($E<10$ meV) in NaFeAs are dominated by $c$-axis 
polarized excitations \cite{Yu13}, while the neutron
spin resonance at $E_{r1}$ has both $a$-axis and $c$-axis polarized spin excitation components \cite{CLZhang14}.

Alternatively, the double resonances may be associated with the gap anisotropy induced by the strong 
orbital-selective superconducting pairing \cite{R.Yu}. In the superconducting phase of the detwinned NaFe$_{0.985}$Co$_{0.015}$As, the splitting between the $d_{xz}$ and $d_{yz}$ orbitals strongly modifies the Fermi surface nesting condition, as shown in 
Fig. 1(c). As a consequence, the superconducting gaps associated with different orbitals are also different, \emph{i.e.}, $\Delta_{xz}\neq\Delta_{yz}\neq\Delta_{xy}$. Since the Fermi surfaces have mixed orbital character, such an orbital dependent pairing will give rise to gap anisotropy along the Fermi surfaces. It will also make the resonance energies very different between the intra-orbital ($d_{yz}$-$d_{yz}$) and inter-orbital ($d_{yz}$-$d_{xy/xz}$) scatterings, although both scatterings may take place at the same wave vector. Therefore, it is expected that these intra- and inter-orbital scatterings with different energy scales lead to two spin resonance peaks at $\mathbf{Q}_1$, just as observed in our experiment.

In conclusion, our INS experiments on partially detwinned NaFe$_{0.985}$Co$_{0.015}$As shows that neutron spin resonance in this system only appear at the AF wave vector $\textbf{Q}_{AF} = (1,0)$.  
We connect our observations with the anisotropic band shifting below $T_s$ in NaFe$_{1-x}$Co$_{x}$As superconductors. The 
$d_{yz}$/$d_{xz}$ orbital degeneration breaks at $T_{s}$ (or higher temperatures under uniaxial pressure), and bands with dominant d$_{yz}$ orbital character shift up in energy and have better nesting conditions \cite{M.Yi,Y.Zhang}. 
Our analysis agrees with such band structure change and indicates neutron spin resonance in NaFe$_{0.985}$Co$_{0.015}$As 
reveals a strong orbital dependent superconducting pairing enhanced by the reconstruction of the band structure below $T_{s}$, in which the scatterings associated with the $d_{yz}$ orbital play a crucial role.  These results suggest that 
intra-orbital quasiparticle scattering of the $d_{yz}$-$d_{yz}$ orbitals are important for superconductivity, similar to 
magnetic scattering of LiFe$_{1-x}$Co$_{x}$As family of materials \cite{L.Yu}.

The authors gratefully acknowledge Dr. Y. Ming and for helpful discussions. The single crystal
growth and neutron scattering work at Rice is supported by the U.S. DOE, BES
under contract no. DE-SC0012311 (P.D.). A part of the material synthesis work at Rice is
supported by the Robert A. Welch Foundation Grant No. C-1839 (P.D.).
R.Y. was supported in part by the National Science Foundation of China Grant No. 11374361 and the Fundamental Research Funds for the Central Universities and the Research Funds of Renmin University of China Grant No. 14XNLF08.

\end{document}